\documentstyle[sprocl,epsf]{article}
\newcommand{\Nf}{N_{\!f}} 
\newcommand{\MSbar}{\overline{\mbox{MS}}} 
\newcommand{\Dslash}{D \! \! \! \! /}

\newcommand{\half}{\mbox{\small{$\frac{1}{2}$}}} 
 
\begin{document}
\begin{flushright} 
{\bf {LTH 383}} 
\end{flushright} 
\vspace{0.4cm} 
\title{PROGRESS WITH LARGE $\Nf$ $\beta$-FUNCTIONS \footnote{Talk 
presented at 5th International Workshop on Software Engineering, Neural Nets, 
Genetic Algorithms, Expert Systems, Symbolic Algebra and Automatic 
Calculations, Lausanne, Switzerland, 2-6th September, 1996.} 
} 
\author{ J.A. GRACEY }
\address{Department of Mathematical Sciences, University of Liverpool, P.O. Box
147, Liverpool, L69 3BX, United Kingdom} 
\maketitle\abstracts{We report on the progress in the computation of the 
$\beta$-functions of $\phi^4$ theory and QCD in the large $N$ expansion. For 
the former we give an analytic formula for the critical exponent which encodes 
higher order coefficients in the series beyond those currently known in the 
$\MSbar$ scheme at $O(1/N^2)$ in a large $N$ expansion. This allows us to 
deduce new coefficients in the field anomalous dimension which relate to 
Kreimer's knot theory of the divergences of a quantum field theory. A similar 
exponent is computed for the $\beta$-function of QCD at $O(1/\Nf)$.}{}{}  
 
\section{Introduction}    
We report on two areas where progress has been made into revealing the higher 
order structure of $\beta$-functions using the large $N$ expansion. First, we 
recall that a highlight of the Pisa workshop was the breakthrough by Broadhurst
in computing those diagrams of the six and seven loop $\beta$-function in 
$\phi^4$ theory which are free of subdivergences.\cite{1} This was in response 
to Kreimer's association of the divergences of multiloop Feynman diagrams 
involving transcendental numbers with knot theory and in particular positive 
braids.\cite{2} For example, the appearance of the Riemann zeta series, 
$\zeta(2n-3)$ $n$ $\geq$ $3$, is well known in high order $\MSbar$ calculations
and these were associated with $(2n - 3,2)$ torus knots.\cite{2} The six and 
seven loop terms of the $\beta$-function contained new transcendentals.\cite{1}
These were represented by double or triple sums which could not be reduced to 
products of $\zeta(n)$. They were associated with the torus knot $10_{124}$ 
and the hyperbolic knots $10_{139}$, $10_{152}$ and $11_{353}$. As these 
interesting knot numbers appear at very high orders it is important to have an 
efficient way of accessing them. The large $N$ approach is a technique which 
facilitates this.\cite{3} For instance in $\phi^4$ theory with an $O(N)$ 
symmetry the field anomalous dimension, $\gamma(g)$, has been computed 
analytically in $d$-dimensions at $O(1/N^3)$ at a fixed point, $g_c$, using a 
conformal bootstrap method.\cite{4} The result involves a particular two loop 
Feynman integral whose $\epsilon$-expansion, where $d$ $=$ $4$ $-$ $2\epsilon$, 
contains irreducible double sums. These therefore can be associated with the 
more complicated positive braids. To find at what order in perturbation theory 
they arise and to determine the explicit $\MSbar$ coefficients in $\gamma(g)$ 
associated with them one needs to solve the critical renormalization group 
relation $\eta$ $=$ $2\gamma(g_c)$. Necessary for this is the location of the 
critical coupling $g_c$ $=$ $g_c(\epsilon,N)$ which is dimensionless in 
$4$-dimensions. In this article we report the result of the $O(1/N^2)$ 
calculation of $\beta^\prime(g_c)$ and therefore $g_c$ which allows us to 
proceed. The remainder of the paper is devoted to a similar calculation of the 
$\beta$-function of QCD,\cite{5} though at $O(1/\Nf)$, which will provide {\em 
new} $\MSbar$ coefficients and therefore an independent check for future $4$ 
and higher loop calculations, albeit at the easier end of the range of 
perturbative calculability.  

\section{$\phi^4$ $\beta$-function} 
The large $N$ method to compute all orders coefficients of the renormalization 
group functions at a fixed point of the $\beta$-function has been discussed 
previously.\cite{6,7} For $\phi^4$ theory with an $O(N)$ symmetry we use the 
lagrangian in the form  
\begin{equation} 
L ~=~ \half (\partial \phi^i)^2 ~+~ \half \sigma \phi^i \phi^i ~-~ \sigma^2/2g 
\end{equation} 
where $\sigma$ is an auxiliary field. This form is used so that uniqueness can 
be applied and also to emphasis the equivalence with the $O(N)$ $\sigma$ model 
in $(2$ $-$ $2\epsilon)$ dimensions.\cite{8} This $d$-dimensional equivalence 
means that the critical exponents computed in either model are the same. 
Although various $\sigma$ model exponents aside from $\eta$ are known at 
$O(1/N^2)$,\cite{6} the exponent $\omega$ which we define as $\omega$ $=$ $-$ 
$\beta^\prime(g_c)/2$ has not been determined. It plays a role in understanding
corrections to scaling and as it is not related through scaling laws it must be 
computed independently. 

In the large $N$ formalism\cite{6} this is achieved by considering corrections
to asymptotic scaling and then solving the $\phi$ and $\sigma$ Schwinger Dyson 
equations self consistently. At leading order due to the structure of the 
$N$-dependence of the $2$ $\times$ $2$ matrix yielding $\omega$ there is one
$2$ loop and one $3$ loop Feynman diagram to compute. This situation is 
parallel to the same computation in the Gross Neveu model.\cite{9} For the 
$O(1/N^2)$ calculation, in addition to computing the $O(1/N)$ corrections to 
the previous graphs, there are two $2$-loop, eight $3$-loop, eleven $4$-loop, 
ten $5$-loop and two $6$-loop integrals which need to be evaluated. We 
find,\cite{3}  
\begin{equation} 
\omega_1 ~=~ (2\mu-1)^2\eta_1    
\end{equation} 
and 
\begin{eqnarray} 
\omega_2 &=& \eta_1^2 \left[ -~   
\frac{4(\mu^2-5\mu+5)(2\mu-3)^2(\mu-1)\mu^2 ( \bar{\Phi} + \bar{\Psi}^2 )} 
{(\mu-2)^3(\mu-3)} \right. \nonumber \\ 
&&-~ \left. \frac{16\mu(2\mu-3)^2}{(\mu-2)^3(\mu-3)^2\eta_1} 
\right. \nonumber \\  
&&-~ \left. \frac{3(4\mu^5-48\mu^4+241\mu^3-549\mu^2+566\mu-216)(\mu-1)\mu^2 
\hat{\Theta}}{2(\mu-2)^3(\mu-3)} \right. \nonumber \\ 
&&-~ \left. [16\mu^{10}-240\mu^9+1608\mu^8-6316\mu^7+15861\mu^6 \right. 
\nonumber \\ 
&&~~~~ \left. -~ 25804\mu^5+26111\mu^4-14508\mu^3+2756\mu^2 \right. 
\nonumber \\
&&~~~~ \left. +~ 672\mu-144)]/[(\mu-2)^4(\mu-3)^2] \bar{\Psi} \right. 
\nonumber \\ 
&&+~ \left. [144\mu^{14}-2816\mu^{13}+24792\mu^{12}-130032\mu^{11} 
+452961\mu^{10} \right. \nonumber \\ 
&&~~~~ \left. -~ 1105060\mu^9+1936168\mu^8-2447910\mu^7+2194071\mu^6 \right. 
\nonumber \\ 
&&~~~~ \left. -~ 1320318\mu^5+460364\mu^4-43444\mu^3-26280\mu^2 \right. 
\nonumber \\ 
&&~~~~ \left. +~ 8208\mu-864]/[2(2\mu-3)(\mu-1)(\mu-2)^5(\mu-3)^2\mu] \frac{}{}
\right] 
\end{eqnarray} 
where $\eta_1$ $=$ $-$ $4\Gamma(2\mu-2)/[\Gamma(2-\mu)\Gamma(\mu-1)\Gamma(\mu-2)
\Gamma(\mu+1)]$, $\bar{\Psi}(\mu)$ $=$ $\psi(2\mu-3)$ $+$ $\psi(3-\mu)$ $-$
$\psi(\mu-1)$ $-$ $\psi(1)$, $\hat{\Theta}(\mu)$ $=$ $\psi^\prime(\mu-1)$ $-$
$\psi^\prime(1)$, $\bar{\Phi}(\mu)$ $=$ $\psi^\prime(2\mu-3)$ $-$ 
$\psi^\prime(3-\mu)$ $-$ $\psi^\prime(\mu-1)$ $+$ $\psi^\prime(1)$ and $\omega$
$=$ $\sum_{i=0}^\infty \omega_i/N^i$ with $\omega_0$ $=$ $\mu$ $-$ $2$. The 
$\epsilon$-expansion of Eq. (3) near $d$ $=$ $4$ $-$ $2\epsilon$ dimensions 
gives total agreement with the coefficients of the explicit $5$-loop $\MSbar$ 
results.\cite{10} So writing the $\beta$-function in the form where only the 
$O(1/N)$ and $O(1/N^2)$ coefficients are included,  
\begin{equation} 
\beta(g) ~=~ \half (d-4)g ~+~ (a_1N+b_1)g^2 ~+~ \sum_{r=2}^\infty 
(a_rN+b_r)N^{r-2}g^{r+1} 
\end{equation}  
we find the {\em new} $\MSbar$ coefficients  
\begin{eqnarray} 
a_6 &=& [29 + 528\zeta(3) - 432\zeta(4)]/1866240 \nonumber \\ 
a_7 &=& [61 + 80\zeta(3) + 1584\zeta(4) - 1728\zeta(5) ]/26873856 \nonumber \\ 
b_6 &=& [ \, -~ 28160\zeta(7) + 95200\zeta(6) - 150336\zeta(5) 
          - 6912\zeta(4)\zeta(3) \nonumber \\ 
&&~ +~ 14112\zeta(4) + 24064\zeta^2(3) + 11880\zeta(3) + 5661]/466560 
\nonumber \\ 
b_7 &=& [\, -~ 7795200\zeta(8) + 32724480\zeta(7) - 43286400\zeta(6) 
          - 3993600\zeta(5)\zeta(3) \nonumber \\ 
&&~ +~ 31998720\zeta(5) - 622080\zeta^2(4) 
          + 8663040\zeta(4)\zeta(3) + 2538432\zeta(4) \nonumber \\ 
&&~ -~ 11381760\zeta^2(3) - 7461168\zeta(3) - 1125439]/403107840 
\end{eqnarray}  

Knowledge of Eq. (3) means that $g_c$ is available at $O(1/N^3)$ to all orders 
in $\epsilon$ and therefore we can proceed in deducing coefficients in the 
series of $\gamma(g)$, in a form similar to Eq. (4). The expression for $\eta$ 
$=$ $2\gamma(g_c)$ at $O(1/N^3)$, however, involves the $\epsilon$-expansion of
a $2$-loop self energy integral whose evaluation in terms of standard 
integration techniques for massless Feynman diagrams such as uniqueness is not 
possible for arbitrary $d$.\cite{4} Using group theory, for example, it has 
recently been written in terms of an ${}_3F_2$ hypergeometric function.\cite{3} 
Therefore, writing the perturbative form of $\gamma(g)$ as 
\begin{equation} 
\gamma(g) ~=~ \sum_{r=1}^\infty (c_rN^2+d_rN+e_r)N^{r-2}g^{r+1} 
\end{equation}  
with $e_1$ $\equiv$ $0$ where we have only indicated the leading three orders
in the $1/N$ expansion, we deduce that the first appearance of a non-zeta 
number at this order in $1/N$ is in $e_8$ or ninth loop. Explicitly  
\begin{eqnarray}  
e_8 &=& [1044119552\zeta(9) - 2854457088\zeta(8) + 2246396928\zeta(7) 
\nonumber \\ 
&&~ +~ 264929280\zeta(6)\zeta(3) + 572052480\zeta(6) + 77856768\zeta(5)\zeta(4)
\nonumber \\ 
&&~ -~ 663797760\zeta(5)\zeta(3) - 887356416\zeta(5) - 127844352\zeta^2(4)
\nonumber \\ 
&&~ +~ 309288960\zeta(4)\zeta(3) - 476930304\zeta(4) - 8978432\zeta^3(3) 
\nonumber \\  
&&~ +~ 286906368\zeta^2(3) + 677644800\zeta(3) \nonumber \\ 
&&~ +~ 12386304U_{62} - 96663009]/216710774784 
\end{eqnarray} 
In Eq. (7) $U_{62}$ is the level $8$ non-zeta number first discovered in 
Ref. [11]. In the $\beta$-function calculation it occurs at sixth loop and is 
associated with the $8_{19}$ $=$ $(4,3)$ torus knot.\cite{1}  

\section{QCD $\beta$-function} 
We now report on a similar calculation\cite{5} for the $\beta$-function of QCD 
but at $O(1/\Nf)$ where $\Nf$ is the number of quark flavours. An essential 
feature of the previous section was the equivalence of $\phi^4$ theory at the 
fixed point to another model. The property of universality meant that exponents
calculated in either model gave the same results. This principle can be 
exploited in the case of QCD. In large $\Nf$ it has been demonstrated, at least
in leading order, that QCD is equivalent to a lower dimensional model.\cite{12}
Therefore it is possible to compute $d$-dimensional critical exponents in 
either model but the coefficients of the perturbative renormalization group 
functions of QCD are deduced from knowledge of the location of the QCD fixed 
point. The main advantage of using this equivalence in this instance is that 
the lower dimensional model has a simpler form. In particular it has fewer 
interactions which immediately reduces the number of Feynman diagrams which 
need to be computed. More concretely the lagrangian for QCD, in standard 
notation, is  
\begin{equation} 
L ~=~ i \bar{\psi}^{iI} \Dslash \psi^{iI} ~-~ \frac{(G^a_{\mu \nu})^2}{4e^2}  
\end{equation} 
The model to which it is equivalent as $\Nf$ $\rightarrow$ $\infty$ is the 
non-abelian Thirring model, (NATM), which is perturbatively renormalizable in
strictly two dimensions and has the lagrangian, 
\begin{equation} 
L ~=~ i \bar{\psi}^{iI} \Dslash \psi^{iI} ~-~ \frac{(A^a_\mu)^2}{2\lambda}  
\end{equation} 
where the coupling constant $\lambda$ is dimensionless in $2$-dimensions. 
Clearly the NATM contains only a single quark gluon interaction. In using Eq. 
(9) to compute exponents there would appear to be no contributions to 
quantities in QCD which are built with the triple and quartic gluon 
self-interactions. It was demonstrated, however,\cite{12} that these pieces are
contained in subdiagrams which involve a quark loop which connects three and
four gluon lines respectively. In other words using the NATM with the critical
point massless propagators diagrams with such loops reproduce the contribution
to the perturbation theory in QCD of both the loop itself and also the gluon 
self-interactions. Indeed this is apparent in the results we have previously 
computed using Eq. (9) in relation to the known explicit $3$-loop $\MSbar$ 
results.\cite{7} Further in dealing with non-abelian gauge theories we use a 
covariant gauge fixing. Therefore ghost fields must also be included in both 
lagrangians. It turns out that since a loop of ghost fields gives one power 
less of $\Nf$ than at leading order in $1/\Nf$ they do not contribute to the 
exponents.  

To find $\omega$ one computes the dimension of the operator associated with 
that coupling.\cite{5} From Eq. (8) this is $G$ $=$ $(G^a_{\mu\nu})^2$ and from 
the last term of Eq. (8) it is easy to deduce the scaling law which relates 
$\omega$ to the anomalous dimension of $G$ and the gluon anomalous dimension. 
As the $O(1/\Nf)$ result for the abelian sector has been reported 
previously\cite{13}$^{\!,\,}$\cite{14} all that needs to be computed are the 
diagrams in the gluon $2$-point function with $G$ inserted whose colour factor 
is proportional to $C_2(G)$. There are two such graphs; one two loop and one 
three loop. Calculating these with the massless critical point propagators we 
obtain the result  
\begin{eqnarray} 
\omega_1 &=& -~ \left[ \frac{}{} (2\mu-3)(\mu-3)C_2(R) \right. \nonumber \\ 
&&~~~~ \left. -~ \frac{(4\mu^4 - 18\mu^3 + 44\mu^2 - 45\mu + 14)C_2(G)} 
{4(2\mu-1)(\mu-1)} \right] \frac{\eta^{\mbox{o}}_1}{T(R)}  
\end{eqnarray} 
This has been computed in an arbitrary covariant gauge and we have checked that
the gauge parameter vanishes as it ought. Except in the Landau gauge, however,
there is potential for the operator $G$ to mix with $(\partial^\mu A^a_\mu)^2$.
It turns out that the mixing matrix at criticality is in fact diagonal. The 
final check we have is in relation to the $\epsilon$-expansion of Eq. (10) and 
we record that it agrees with the $3$-loop $\MSbar$ QCD 
$\beta$-function.\cite{15,16,17} This agreement further justifies our previous 
remarks concerning the use of the NATM in computing exponents in QCD. In 
similar notation to Eq. (4) we deduce the {\em new} higher order coefficients  
\begin{eqnarray} 
a_4 &=& - ~ [154C_2(R) + 53C_2(G)]/3888 \nonumber \\ 
a_5 &=& [(288\zeta(3) + 214)C_2(R) + (480\zeta(3) - 229)C_2(G)]/31104 
\nonumber \\ 
a_6 &=& [(864\zeta(4) - 1056\zeta(3) + 502)C_2(R) \nonumber \\
&&~ +~ (1440\zeta(4) - 1264\zeta(3) - 453)C_2(G)]/233280 \nonumber \\  
a_7 &=& [(3456\zeta(5) - 3168\zeta(4) - 2464\zeta(3) + 1206)C_2(R) \nonumber \\
&&~ +~ (5760\zeta(5) - 3792\zeta(4) - 848\zeta(3) - 885)C_2(G)]/1679616  
\end{eqnarray} 

\section{Conclusion} 
We have provided the two leading $\MSbar$ numbers in the quintic in $N$ which 
appears as the coefficient in the $\beta$-function of $\phi^4$ theory at six
loops. Although information on the hardest diagrams has been determined in the 
$N$ $=$ $1$ case,\cite{1,2} computation of the remaining numbers to complete 
the six loop term is still far off. For this to be achieved a very high degree 
of ingenuity will be required to produce an algorithm which can then be 
implemented in some symbolic manipulation programme. 

\section*{Acknowledgements}
The work reported in this article has been carried out in part with the support
of a {\sc pparc} Advanced Fellowship. The calculations were performed with the
use of {\sc Reduce} version 3.4 and {\sc Form} version 2.2c.  
 

\end{document}